\documentclass[twocolumn,english,showpacs,twocolumn,superscriptaddress,nofootinbib,aps]{revtex4-1}
\usepackage[T1]{fontenc}
\usepackage[latin9]{inputenc}
\usepackage{babel}
\usepackage{amsmath}
\usepackage{graphicx}
\usepackage[unicode=true,pdfusetitle,
 bookmarks=true,bookmarksnumbered=false,bookmarksopen=false,
 breaklinks=false,pdfborder={0 0 1},backref=false,colorlinks=false]
 {hyperref}
\begin{document}
\title{Signals of quark combination at hadronization in $pp$ collisions
at $\sqrt{s}=200$ GeV }
\author{Jun Song }
\affiliation{School of Physical Science and Intelligent Engineering, Jining University,
Shandong 273155, China}
\author{Hai-hong Li}
\affiliation{School of Physical Science and Intelligent Engineering, Jining University,
Shandong 273155, China}
\author{Feng-lan Shao}
\email{shaofl@mail.sdu.edu.cn}

\affiliation{School of Physics and Physical Engineering, Qufu Normal University,
Shandong 273165, China}
\begin{abstract}
We find signals of quark combination at hadronization from the experimental
data of $p_{T}$ spectra of hadrons at mid-rapidity in $pp$ collisions
at $\sqrt{s}=200$ GeV. The first is the constituent quark number
scaling property for $p_{T}$ spectra of $\Omega^{-}$ and $\phi$
and that for $p_{T}$ spectra of $p$ and $\rho^{0}$. The second
is that $p_{T}$ spectra of $\Lambda$, $\Xi^{-}$, and $K^{*0}$
can be self-consistently described using the spectrum of strange quarks
from $\phi$ data and that of up/down quarks from $p$ data in the
equal-velocity combination mechanism. The third is that experimental
data for $p_{T}$ spectrum of $D^{*+}$ are also well described using
the spectrum of up/down quarks from $p$ data and that of charm quarks
from perturbative QCD calculations. These results indicate a similarity
between hadron production in $pp$ collisions at $\sqrt{s}=200$ GeV
and that at LHC energies. We predict $p_{T}$ spectra of single-charm
hadrons and their spectrum ratios. We suggest systematic measurements
in $pp$ collisions at $\sqrt{s}=200$ GeV in future so as to better
understand the property of small parton system created in $pp$ collisions
at different collision energies.
\end{abstract}
\maketitle

\section{\emph{Introduction}\label{sec:Intro} }

Hadronization refers to the process of the formation of hadrons from
final state quarks and gluons created in high energy reactions. Hadronization
is a non-perturbative quantum chromodynamics (QCD) process and is
described by phenomenological models at present. String fragmentation
\citep{Andersson:1983ia}, cluster fragmentation \citep{Webber:1983if}
and quark recombination \citep{Bjorken:1973mh,Das:1977cp} are three
kinds of popular models which are often used to describe the hadron
production in high energy reactions. 

Experimental data of hadron production in high energy reactions often
provide new inspiration on the understanding of hadronization. We
recall that heavy-ion collision experiments at Relativistic Heavy
Ion Collider (RHIC) in the 2000s found several surprising phenomena
such as the enhanced ratio of baryon to meson \citep{PHENIX:2001vgc,STAR:2006uve,STAR:2006pcq}
and number-of-constituent quark scaling (NCQ) property for hadronic
elliptic flow \citep{STAR:2003wqp,STAR:2004jwm,PHENIX:2006dpn} in
intermediate $p_{T}$ range. These observations prompt the study of
quark (re-)combination or parton coalescence mechanism \citep{Greco:2003xt,Fries:2003vb,Hwa:2002tu,Molnar:2003ff,Chen:2006vc,Shao:2009uk}
for the hadronization of bulk quark matter created in relativistic
heavy-ion collisions. On the other hand, the hadronization of parton
jet with high $p_{T}$ and small parton system is still usually described
by fragmentation mechanism. 

In the last decade, experiments of $pp$ and pA collisions at energies
available at Large Hadron Collider (LHC) found a series of new phenomena
in hadron production in high multiplicity events such as ridge or
long-range correlation \citep{Khachatryan:2010gv,CMS:2012qk}, collectivity
\citep{Khachatryan:2015waa,Khachatryan:2016txc}, enhanced ratio of
baryon to meson \citep{ALICE:2017jyt,Abelev:2013haa,Adam:2016dau,Adam:2015vsf}.
These phenomena have been observed in relativistic heavy-ion collisions
and are usually regarded to be closely related to the formation of
quark-gluon plasma (QGP). Observation of these phenomena in $pp$
and $p$A collisions therefore invoke an interesting question, i.e.,
the possible creation of mini-QGP. This attracts intensive theoretical
studies from different aspects \citep{Liu:2011dk,Bzdak:2013zma,Bozek:2013uha,Prasad:2009bx,Avsar:2010rf,Zhao:2017rgg}.
Our recent studies \citep{Shao:2017eok,Song:2017gcz,Gou:2017foe,Zhang:2018vyr,Li:2017zuj,Song:2018tpv,Li:2021nhq}
found that an equal-velocity combination (EVC) mechanism of constituent
quarks and antiquarks can systematically describe $p_{T}$ spectra
of light-flavor and single-charm hadrons. Compared with the traditional
viewpoint that fragmentation mechanism is often applied to small parton
system and usually successful, our studies indicate the new feature
of hadron production in $pp$ and $p$A collisions at LHC energies.
This may be related to the possible formation of mini-QGP in $pp$
and $p$A collisions at LHC energies.

The production of identified hadrons in $pp$ collisions at $\sqrt{s}=$
200 GeV was systematically measured by STAR collaboration in early
years of RHIC experiments \citep{STAR:2004yym,STAR:2003vqj,STAR:2004bgh,STAR:2006nmo,STAR:2006xud,STAR:2012nbd,Qiu:2016tie}.
Experimental data were usually compared with calculations of event
generators such as PHTHIA with tuned parameters. In view of our findings
in $pp$ collisions at LHC energies \citep{Shao:2017eok,Song:2017gcz,Gou:2017foe,Zhang:2018vyr,Li:2017zuj,Song:2018tpv,Li:2021nhq},
it is interesting to study the performance of quark combination in
$pp$ collisions at $\sqrt{s}=$200 GeV so as to find the similarity
or difference in hadron production in $pp$ collisions at two collision
energy scales. The study of $p_{T}$ spectra of identified hadrons
in this paper gives a surprising indication. 

\section{\emph{Quark number scaling of hadronic $p_{T}$ spectra} \label{sec:qns} }

In our EVC model \citep{Song:2017gcz,Gou:2017foe}, a hadron is formed
by the combination of (anti-)quarks with the equal velocity. $p_{T}$
distribution of a hadron ($dN/dp_{T}$) is the product of those of
(anti-)quarks 
\begin{align}
f_{B_{i}}\left(p_{T}\right) & =\kappa_{B_{i}}f_{q_{1}}\left(x_{1}p_{T}\right)f_{q_{2}}\left(x_{2}p_{T}\right)f_{q_{3}}\left(x_{3}p_{T}\right),\label{eq:fbi}\\
f_{M_{i}}\left(p_{T}\right) & =\kappa_{M_{i}}f_{q_{1}}\left(x_{1}p_{T}\right)f_{\bar{q}_{2}}\left(x_{2}p_{T}\right).\label{eq:fmi}
\end{align}
Here, (anti-)quarks are constituent (anti-)quarks so that their equal
velocity combination can correctly construct the on-shell hadron.
Moment fractions satisfy $x_{1}+x_{2}+x_{3}=1$ with $x_{i}=m_{i}/(m_{1}+m_{2}+m_{3})$
($i=1,2,3)$ in baryon formation and $x_{1}+x_{2}=1$ with $x_{i}=m_{i}/(m_{1}+m_{2})$
($i=1,2)$ in meson formation. $m_{i}$ is constituent mass of quark
$q_{i}$. Coefficients $\kappa_{B_{i}}$ and $\kappa_{M_{i}}$ are
independent of $p_{T}$ but dependent on numbers of quarks and antiquarks
\citep{Gou:2017foe}. 

For hyperon $\Omega^{-}(sss)$ which only consists of strange quarks,
its $p_{T}$ distribution has a simple expression 
\begin{align}
f_{\Omega}\left(p_{T}\right) & =\kappa_{\Omega}\left[f_{s}\left(p_{T}/3\right)\right]^{3}.\label{eq:fpt_Omega}
\end{align}
$p_{T}$ distribution of meson $\phi(s\bar{s})$ also has a simple
expression
\begin{equation}
f_{\phi}\left(p_{T}\right)=\kappa_{\phi}f_{s}\left(p_{T}/2\right)f_{\bar{s}}\left(p_{T}/2\right)=\kappa_{\phi}\left[f_{s}\left(p_{T}/2\right)\right]^{2},\label{eq:fpt_phi}
\end{equation}
where the approximation $f_{s}\left(p_{T}\right)=f_{\bar{s}}\left(p_{T}\right)$
at mid-rapidity is taken. From Eqs.~(\ref{eq:fpt_Omega}) and (\ref{eq:fpt_phi}),
we obtain a relationship 
\begin{equation}
f_{\phi}^{1/2}\left(2p_{T}\right)=\kappa_{\phi,\Omega}f_{\Omega}^{1/3}\left(3p_{T}\right)\label{eq:qns_s}
\end{equation}
which is called the constituent quark number scaling of hadronic $p_{T}$
spectra. Coefficient $\kappa_{\phi,\Omega}=\kappa_{\phi}^{1/2}/\kappa_{\Omega}^{1/3}$
is independent of $p_{T}$. For $p_{T}$ spectra of proton and $\rho$,
we obtain a similar relationship
\begin{equation}
f_{\rho}^{1/2}\left(2p_{T}\right)=\kappa_{\rho,p}f_{p}^{1/3}\left(3p_{T}\right)\label{eq:qns_u}
\end{equation}
where approximations $f_{u}(p_{T})=f_{d}(p_{T})$ and $f_{u}\left(p_{T}\right)=f_{\bar{u}}\left(p_{T}\right)$
at mid-rapidity are taken. We run PYTHIA 8 with default parameter
values and find that calculation results do not exhibit properties
in Eqs. (\ref{eq:qns_s}) and (\ref{eq:qns_u}). 

\begin{figure}[h]
\begin{centering}
\includegraphics[width=0.5\linewidth]{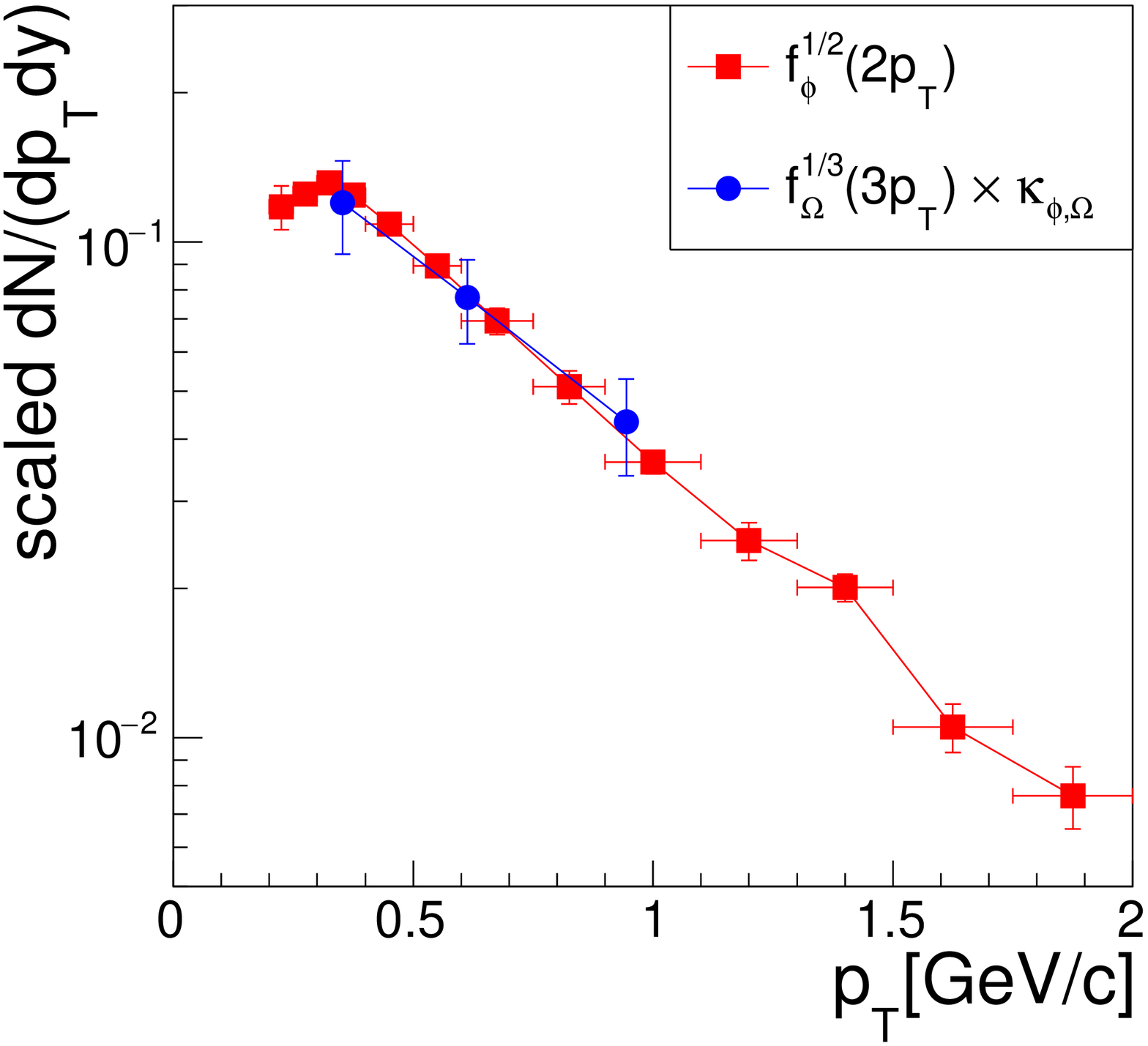}\includegraphics[width=0.5\linewidth]{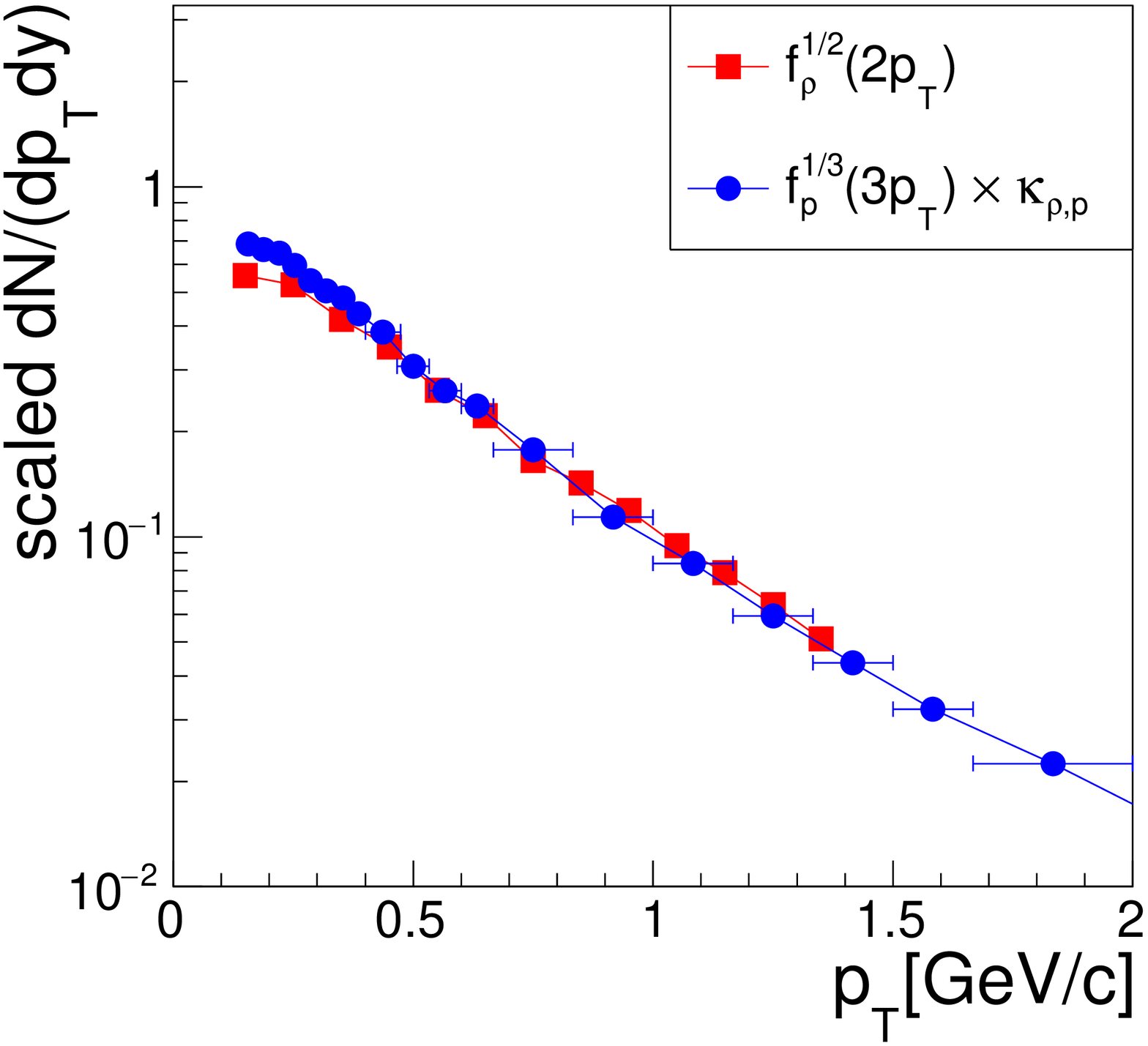}\caption{Test of quark number scaling property for $p_{T}$ spectra of hadrons
in $pp$ collisions at $\sqrt{s}=$200 GeV by using experimental data
of STAR collaboration \citep{STAR:2004yym,STAR:2006xud,STAR:2006nmo,STAR:2003vqj}.
$\kappa_{\phi,\Omega}$ is 1.88 and $\kappa_{\rho,p}$ is 1.10.\textbf{
}The line between datum points is for eye guide of the shape of the
scaled data. \label{fig:qns}}
\par\end{centering}
\end{figure}

 In Fig.~\ref{fig:qns} (a), we test the scaling property Eq.~(\ref{eq:qns_s})
using experimental data of $\Omega^{-}+\bar{\Omega}^{+}$ and $\phi$
at mid-rapidity in inelastic $pp$ collisions at $\sqrt{s}=$200 GeV
\citep{STAR:2004yym,STAR:2006nmo}. $\kappa_{\phi,\Omega}$ is taken
as 1.88. $\Omega^{-}+\bar{\Omega}^{+}$ has only three datum points
and we see that they are almost coincident with the scaled data of
$\phi$. In Fig.~\ref{fig:qns} (b), we test Eq.~(\ref{eq:qns_u})
using experimental data of proton and $\rho^{0}$ \citep{STAR:2006xud,STAR:2003vqj}.
$\kappa_{\rho,p}$ is taken as 1.10. Except for the first datum point
at $p_{T,u}\approx0.15$ GeV/c, we see that other datum points of
$\rho^{0}$ are very close to the scaled data of proton. We emphasize
that values of two coefficients $\kappa_{\phi,\Omega}$ and $\kappa_{\rho,p}$
can be reproduced in our model by considering quark number distributions
at hadronization. Therefore, these two scaling tests positively indicate
quark combination mechanism at hadronization in $pp$ collisions even
at $\sqrt{s}=200$ GeV. 

\section{\emph{$p_{T}$ spectra of mixing-flavor hadrons} \label{sec:fpt_LamXi} }

Subsequently, we understand the experimental data for $p_{T}$ spectra
of $\Lambda$, $\Xi^{-}$ and $K^{*0}$ \citep{STAR:2004bgh,STAR:2006nmo}.
These hadrons consist of strange quarks and up/down quarks. By Eqs.
(\ref{eq:fbi}) and (\ref{eq:fmi}), their $p_{T}$ spectra at hadronization
are given as 
\begin{align}
f_{\Lambda}\left(p_{T}\right) & =\kappa_{\Lambda}\left[f_{u}\left(\frac{1}{2+r_{su}}p_{T}\right)\right]^{2}f_{s}\left(\frac{r_{su}}{2+r_{su}}p_{T}\right),\label{eq:fpt_Lam}\\
f_{\Xi}\left(p_{T}\right) & =\kappa_{\Xi}\left[f_{s}\left(\frac{r_{su}}{1+2r_{su}}p_{T}\right)\right]^{2}f_{u}\left(\frac{1}{1+2r_{su}}p_{T}\right),\label{eq:fpt_Xi}\\
f_{K^{*}}\left(p_{T}\right) & =\kappa_{K^{*}}f_{u}\left(\frac{1}{1+r_{su}}p_{T}\right)f_{s}\left(\frac{r_{su}}{1+r_{su}}p_{T}\right)\label{eq:fpt_Kstar}
\end{align}
where $r_{su}=m_{s}/m_{u}$ is the relative momentum ratio of strange
quark to up quark. We take $r_{su}=1.67$ by considering constituent
quark masses $m_{s}=0.5\sim0.55$ GeV and $m_{u}=0.3\sim0.33$ GeV
in constituent quark model. To calculate Eqs.~(\ref{eq:fpt_Lam})-(\ref{eq:fpt_Kstar}),
quark distributions $f_{s}(p_{T})$ and $f_{u}(p_{T})$ at hadronization
are needed. We obtain them by using our EVC model to fit experimental
data of $\phi$ and proton \citep{STAR:2004yym,STAR:2006nmo}. Here,
the decay contributions of decuplet baryons in ground state to octet
baryons are included. The detailed derivation of coefficient $\kappa_{h}$
in the EVC model can be found in Refs. \citep{Gou:2017foe,Zhang:2018vyr,Li:2021nhq}.

\begin{figure}[h]

\begin{centering}
\includegraphics[width=0.95\linewidth]{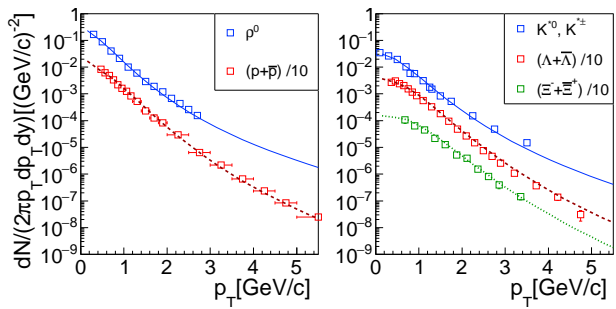}\caption{$p_{T}$ spectra of hadrons at mid-rapidity in $pp$ collisions at
$\sqrt{s}=200$ GeV. Lines are model results and symbols are experimental
data \citep{STAR:2004bgh,STAR:2006nmo}. \label{fig:fpt_lh}}
\par\end{centering}
\end{figure}

In Fig.~\ref{fig:fpt_lh} (a), we firstly show $p_{T}$ spectrum
of $\rho^{0}$ based on $f_{u}(p_{T})$ fitted from proton data. We
see that $\rho^{0}$ result is in good agreement with experimental
data \citep{STAR:2003vqj}. We note that the consistency between $\rho^{0}$
and proton here is better than the scaling test in Fig. \ref{fig:qns}(b).
This is because final-state protons receive certain decay contamination
of decuplet baryons $\Delta$, which will weakly influence $p_{T}$
spectrum of proton. $\Omega$ and $\phi$ hardly contain decay contributions
and therefore their $p_{T}$ spectra do not have this contamination. 

In Fig.~\ref{fig:fpt_lh}(b), we show results for $p_{T}$ spectra
of $\Lambda$, $\Xi^{-}$ and $K^{*0}$. We see a good agreement with
experimental data of three hadrons \citep{STAR:2004bgh,STAR:2006nmo}.
Combining results of Figs.~\ref{fig:qns} and \ref{fig:fpt_lh},
we see that experimental data of $\phi$, $\Omega^{-}$, $\rho^{0}$,
proton, $\Lambda$, $\Xi^{-}$ and $K^{*0}$ can be self-consistently
explained by a set of quark spectra at hadronization $f_{u}(p_{T})$
and $f_{s}(p_{T})$ under equal-velocity combination mechanism. This
is the explicit signal of quark combination at hadronization in $pp$
collisions at $\sqrt{s}=200$ GeV. 

\section{\emph{$p_{T}$ spectrum of single-charm hadron }$D^{*+}$ \label{sec:fc} }

We extend the above study to the combination of charm quark and light-flavor
quarks. Because constituent mass of charm quark is larger than those
of light-flavor quarks, a charm quark with momentum $p_{T}$ will
hadronize by combining a light-flavor antiquark or two light-flavor
quarks with momentum $p_{T}/r_{cl}$ where $r_{cl}=m_{c}/m_{l}$ ($l=u,s$).
We take $r_{cu}=5$ and $r_{cs}=3$ by considering the constituent
mass of charm quark $m_{c}=1.5\sim1.7$ GeV. In our EVC model, $p_{T}$
distribution of $D^{*+}$ is 
\begin{equation}
f_{D^{*}}\left(p_{T}\right)=\kappa_{D^{*}}f_{c}\left(\frac{r_{cu}}{1+r_{cu}}p_{T}\right)f_{u}\left(\frac{1}{1+r_{cu}}p_{T}\right)
\end{equation}
where we assume $f_{u}\left(p_{T}\right)=f_{\bar{u}}\left(p_{T}\right)$
at mid-rapidity. 

\begin{figure}[h]
\centering{}\includegraphics[width=0.95\linewidth]{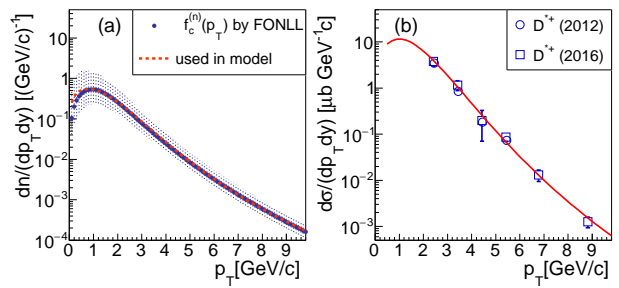}\caption{The charm quark distribution at hadronization (a) and result for differential
cross-section of $D^{*+}$ (b) in $pp$ collisions at $\sqrt{s}=$200
GeV. Symbols in panel (b) are experimental data of $D^{*+}$ \citep{STAR:2012nbd,Qiu:2016tie}.\label{fig:fc} }
\end{figure}

Since $f_{u}\left(p_{T}\right)$ is already known by fitting data
of proton, $p_{T}$ spectrum of $D^{*+}$ can be calculated when $f_{c}\left(p_{T}\right)$
is also known. Here, we consider the calculation result of perturbative
QCD for differential cross-section of charm quark in FONLL scheme
\citep{Cacciari:1998it,Cacciari:2001td}. Because FONLL calculation
has relatively large uncertainties at low $p_{T}$, we firstly fit
the FONLL calculation in Fig. \ref{fig:fc}(a) with a L\'evy-Tsallis
function to get the normalized distribution $f_{c}^{(n)}\left(p_{T}\right)$
and then take $d\sigma_{c}/dy=0.125$ mb at mid-rapidity which is
located in the range of theoretical uncertainties. 

In Fig. \ref{fig:fc}(b), we show model result of differential cross-section
of $D^{*+}$ and compare it with available experimental data \citep{STAR:2012nbd,Qiu:2016tie}.
We see a good agreement. This provides a significant indication on
the equal velocity combination of charm quark with light-flavor (anti-)quarks
as an effective hadronization mechanism in $pp$ collisions at $\sqrt{s}=$200
GeV. 

\section{\emph{Prediction of single-charm hadrons} \label{sec:prediction} }

\begin{figure}[h]
\centering{}\includegraphics[width=0.99\linewidth]{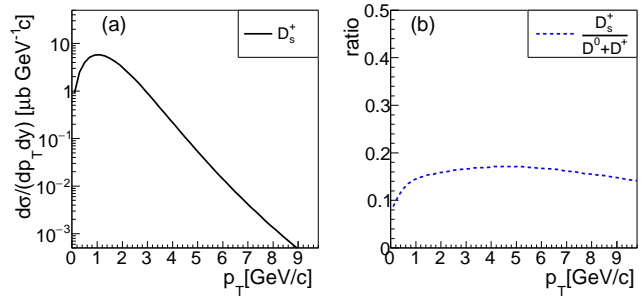}\caption{Differential cross-section of $D_{s}^{+}$ (a) and the spectrum ratio
$D_{s}^{+}/\left(D^{0}+D^{+}\right)$ (b) in $pp$ collisions at $\sqrt{s}=$200
GeV. \label{fig:fpt_Ds}}
\end{figure}

Similarly, we study the combination of charm quark with a strange
antiquark to form a $D_{s}^{+}$. The calculation results for differential
cross-section of $D_{s}^{+}$ and the spectrum ratio $D_{s}^{+}/\left(D^{0}+D^{+}\right)$
as the function of $p_{T}$ are shown in Fig. \ref{fig:fpt_Ds}. Compared
with $D^{0,+}$, production of $D_{s}^{+}$ is suppressed. As we known,
in the light-flavor background the number of strange (anti-)quarks
is smaller than that of up/down (anti-)quarks. Therefore, a charm
has a relatively small chance to capture a co-moving $\bar{s}$ to
form a $D_{s}^{+}$. We use a suppression factor $\lambda_{s}=N_{s}/N_{\bar{u}}$
to denote the relative abundance of strange quarks. In our model yield
ratio of $D_{s}^{+}/\left(D^{0}+D^{+}\right)$ has a simple expression

\begin{equation}
\frac{d\sigma_{D_{s}^{+}}/dy}{d\sigma_{D^{0}+D^{+}}/dy}=\frac{1}{2}\lambda_{s}.
\end{equation}
Since $\lambda_{s}\approx0.29$ in $pp$ collisions at $\sqrt{s}=$200
GeV, we see in Fig. \ref{fig:fpt_Ds}(b) that the spectrum ratio $D_{s}^{+}/\left(D^{0}+D^{+}\right)$
is located in the range {[}0.1,0.2{]}. The ratio has a weak $p_{T}$
dependence, which is because relative abundance of strange quarks
is $p_{T}$ dependent and combination kinematics is slightly different
for $c\bar{s}$ and $c\bar{u}$ pairs. 

\begin{figure}[h]
\centering{}\includegraphics[width=0.99\linewidth]{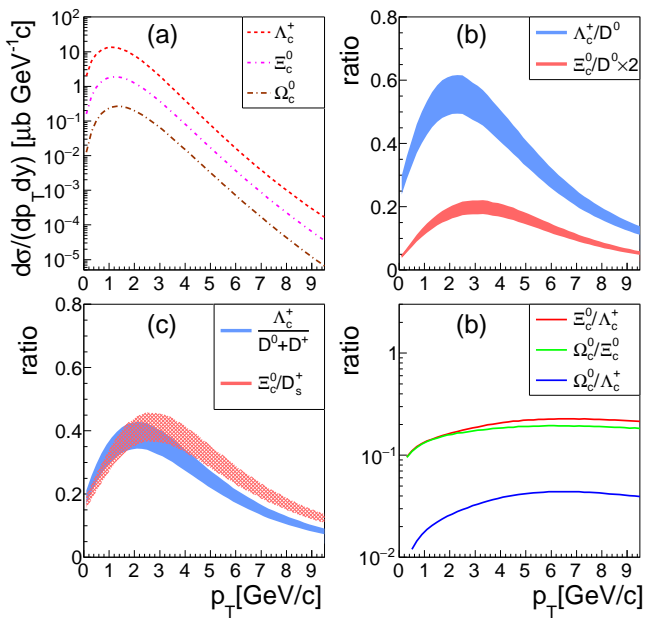}\caption{Differential cross-sections of $\Lambda_{c}^{+}$, $\Xi_{c}^{0}$
and $\Omega_{c}^{0}$ (a) and several ratios among charmed hadrons
in $pp$ collisions at $\sqrt{s}=$200 GeV. \label{fig:fpt_bc}}
\end{figure}

We further calculate $p_{T}$ spectra of single-charm baryons by the
equal-velocity combination of a charm and two light-flavor quarks.
In Fig. \ref{fig:fpt_bc} (a), we present results for differential
cross-sections of $\Lambda_{c}^{+}$, $\Xi_{c}^{0}$ and $\Omega_{c}^{0}$
as the model parameter $R_{B/M}^{(c)}$ is taken as $0.374\pm0.042$.
In quark combination mechanism, a charm can form a meson by picking
up an antiquark or form a baryon by picking up two quarks. Since hadronization
unitarity requires that a charm quark has to become a hadron at last,
there exists a competition between baryon formation and meson formation.
In our model, such a non-perturbative competition dynamic is parameterized
by $R_{B/M}^{(c)}$ and is tuned by experimental data. We fit the
latest experimental data of $\Lambda_{c}^{+}/D^{0}$ in $pp$ collisions
at LHC energies \citep{Acharya:2020lrg,ALICE:2021rzj} and obtain
$R_{B/M}^{(c)}\approx0.374\pm0.042$. Then, we use it to predict the
production of single-charm baryons in $pp$ collisions at $\sqrt{s}=$200
GeV. 

In Fig.~\ref{fig:fpt_bc}(b), we present spectrum ratios $\Lambda_{c}^{+}/D^{0}$
and $\Xi_{c}^{0}/D^{0}$ as the function of $p_{T}$. Two ratios increase
at low $p_{T}$, saturate at $p_{T}\approx2-3$ GeV/c and decrease
at larger $p_{T}$. We emphasize that this non-monotonic $p_{T}$
dependence is a typical signal of our model and is mainly caused by
kinematics of equal-velocity quark combination and the property of
$p_{T}$ spectra of light-flavor quarks. 

In Fig.~\ref{fig:fpt_bc}(c), we present spectrum ratios $\Lambda_{c}^{+}/(D^{0}+D^{+})$
and $\Xi_{c}^{0}/D_{s}^{+}$ in order to better quantify the baryon
to meson production competition for charm quark hadronization. Since
$D^{0}+D^{+}$ including strong and electromagnetic decays contains
all $c\bar{u}$ and $c\bar{d}$ combination channels and $\Lambda_{c}^{+}$
contains all $cuu$, $cud$ and $cdd$ combination channels, the yield
ratio $\Lambda_{c}^{+}/(D^{0}+D^{+})$ directly relates to $R_{B/M}^{(c)}$
in the model 
\begin{equation}
\frac{d\sigma_{\Lambda_{c}^{+}}/dy}{d\sigma_{D^{0}+D^{+}}/dy}=\frac{2}{2+\lambda_{s}}R_{B/M}^{(c)}.
\end{equation}
Strangeness suppression factor $\lambda_{s}$ changes weakly ($0.25\sim0.35$)
in $pp$ collisions and causes little contamination on the ratio.
Therefore, the ratio $\Lambda_{c}^{+}/(D^{0}+D^{+})$ is a sensitive
probe of the relative probability of $cl_{1}l_{2}$ combination against
$c\bar{l}$ combination (here, $l=u,d$) at charm quark hadronization.
Similarly, $\Xi_{c}^{0}(cds)/D_{s}^{+}(c\bar{s})$ denotes the relative
probability of $cds$ combination against $c\bar{s}$ combination.
Since the suppression influence of strange quark is canceled in the
ratio, we have $\Xi_{c}^{0}/D_{s}^{+}=\Lambda_{c}^{+}/(D^{0}+D^{+})$
for yield ratios. In Fig.~\ref{fig:fpt_bc}(c), we also see that
the spectrum ratios $\Lambda_{c}^{+}/(D^{0}+D^{+})$ and $\Xi_{c}^{0}/D_{s}^{+}$
have the same magnitude. The small difference in $p_{T}$ dependence
between two ratios is caused by the combination kinematics, i.e.,
momentum fractions $x_{u}$ and $x_{s}$ are different in combination
with charm quark. 

In Fig.~\ref{fig:fpt_bc}(d), we present ratios $\Xi_{c}^{0}/\Lambda_{c}^{+}$,
$\Omega_{c}^{0}/\Lambda_{c}^{+}$ and $\Omega_{c}^{0}/\Xi_{c}^{0}$
as the function of $p_{T}$. In our model, they are related to the
combination dynamics of increasing number of strange quarks involving
the combination process. Statistical combination symmetry is mainly
used in model and gives in yield ratios
\begin{align}
\frac{d\sigma_{\Xi_{c}^{0}}/dy}{d\sigma_{\Lambda_{c}^{+}}/dy} & =\frac{d\sigma_{\Omega_{c}^{0}}/dy}{d\sigma_{\Xi_{c}^{0}}/dy}=\frac{1}{2}\lambda_{s},\\
\frac{d\sigma_{\Omega_{c}^{0}}/dy}{d\sigma_{\Lambda_{c}^{+}}/dy} & =\frac{1}{4}\lambda_{s}^{2}
\end{align}
where $\lambda_{s}\approx0.29$ in $pp$ collisions at $\sqrt{s}=$200
GeV. We clearly see this flavor hierarchy property in spectrum ratios
in Fig. \ref{fig:fpt_bc}(d). In addition, we see a $p_{T}$ dependence
for three ratios, which is because the difference between $p_{T}$
spectrum of up/down quarks and that of strange quarks at hadronization.

\section{\emph{Summary and discussions \label{sec: summary} }}

In summary, we have applied an equal-velocity quark combination model
to understand the early RHIC data for $p_{T}$ spectra of hadrons
in $pp$ collisions at $\sqrt{s}=$ 200 GeV. We found explicit signals
of quark combination at hadronization. First, we observed a constituent
quark number scaling property for $p_{T}$ spectra of $\Omega$ and
$\phi$ and that of proton and $\rho$. Second, based on the $p_{T}$
spectrum of up/down quarks extracted from proton data and that of
strange quarks extracted from $\phi$ data, we found that data for
$p_{T}$ spectra of $\Lambda$, $\Xi^{-}$ and $K^{*0}$ are also
well described. Third, based on the obtained spectrum of up/down quarks
and that of charm quarks from perturbative QCD calculations, we found
that experimental data for differential cross-section of $D^{*+}$
are also well described. 

Because these properties of hadron production are already found in
$pp$ collisions at LHC energies \citep{Gou:2017foe,Song:2017gcz,Song:2018tpv,Zhang:2018vyr,Li:2021nhq},
the current study indicates a significant similarity between the hadron
production in $pp$ collisions at $\sqrt{s}=$200 GeV and that at
LHC energies. As we known, at LHC energies, some experimental phenomena
such as ridge/long-range correlation \citep{Khachatryan:2010gv,CMS:2012qk},
collectivity \citep{Khachatryan:2015waa,Khachatryan:2016txc}, enhanced
baryon-to-meson ratio \citep{ALICE:2017jyt,Abelev:2013haa,Adam:2016dau,Adam:2015vsf}
were observed as the indication of possible formation of mini-QGP
in $pp$ collisions in high-multiplicity events.  On the other hand,
compared with fragmentation mechanism, quark combination mechanism
is conceptually more suitable to describe the hadronization of QGP
and actually works well in relativistic heavy-ion collisions. Interestingly,
our recent works \citep{Gou:2017foe,Song:2017gcz,Li:2017zuj,Song:2018tpv,Zhang:2018vyr,Li:2021nhq}
suggest that an equal-velocity quark combination mechanism at hadronization
can systematically describe the momentum spectra of hadrons in $pp$
collisions at LHC energies. Therefore, signals of quark combination
found in $pp$ collisions at $\sqrt{s}=$200 GeV in current study,
which indicate the stochastic combination of quarks and antiquarks
at hadronization,  inspire us to consider the possibility of mini-QGP
creation in $pp$ collisions at RHIC energies!

We therefore suggest the systematic measurements in $pp$ collisions
at $\sqrt{s}=$200 GeV in future. These measurements should include
ridge/long-range correlation, collectivity, multiplicity dependence
of hadron production and so on. By a systematic comparison with available
LHC data, these measurements will greatly improve our understanding
for the property of small parton system created in $pp$ collisions
at different collision energies. 

\section{Acknowledgments}

We thank Z.~B. Xu for helpful discussions. This work is supported
in part by Shandong Provincial Natural Science Foundation (ZR2019YQ06,
ZR2019MA053), the National Natural Science Foundation of China under
Grant No. 11975011, and Higher Educational Youth Innovation Science
and Technology Program of Shandong Province (2019KJJ010).

\bibliographystyle{apsrev4-1}
\bibliography{ref}

\end{document}